\documentclass[epj,referee]{svjour}
\usepackage{graphics}

\begin{document}

\title{Phase Diagram of the 3D Bimodal Random-Field Ising Model}
\author{N.G. Fytas\thanks{e-mail: nfytas@phys.uoa.gr} \and A. Malakis\thanks{e-mail: amalakis@phys.uoa.gr}
}
\institute{Department of Physics, Section of Solid State Physics,
University of Athens, Panepistimiopolis, GR 15784 Zografos,
Athens, Greece}

\date{Received: date / Revised version: date}

\abstract{The one-parametric Wang-Landau (WL) method is
implemented together with an extrapolation scheme to yield
approximations of the two-dimensional (exchange-energy,
field-energy) density of states (DOS) of the 3D bimodal
random-field Ising model (RFIM). The present approach generalizes
our earlier WL implementations, by handling the final stage of the
WL process as an entropic sampling scheme, appropriate for the
recording of the required two-parametric histograms. We test the
accuracy of the proposed extrapolation scheme and then apply it to
study the size-shift behavior of the phase diagram of the 3D
bimodal RFIM. We present a finite-size converging approach and a
well-behaved sequence of estimates for the critical disorder
strength. Their asymptotic shift-behavior yields the critical
disorder strength and the associated correlation length's
exponent, in agreement with previous estimates from ground-state
studies of the model.
}
\authorrunning{N.G. Fytas and A. Malakis} \titlerunning{Phase Diagram of the 3D Bimodal Random-Field Ising Model}

\maketitle

\section{Introduction}
\label{sec:1}

The
RFIM~\cite{imry75,aharony76,young77,parisi78,grinstein82,imbrie84,villain84,schwartz85,bray85,houghton85,fisher86,binder86,ogielski86,bricmont87,sethna92}
has been extensively studied both because of its interest as a
simple frustrated system and because of its relevance to
experiments~\cite{belanger83,wong83,birgeneau85,nattermann88,belanger98,birgeneau98}.
The Hamiltonian describing the model is
\begin{equation}
\label{eq:1}
\mathcal{H}=-J\sum_{<i,j>}S_{i}S_{j}-h\sum_{i}h_{i}S_{i},
\end{equation}
where $S_{i}$ are Ising spins, $J>0$ is the nearest-neighbors
ferromagnetic interaction, and $h_{i}$ are independent quenched
random-fields (RF's) obtained here from a bimodal distribution of
the form
\begin{equation}
\label{eq:2}
P(h_{i})=\frac{1}{2}[\delta(h_{i}-1)+\delta(h_{i}+1)].
\end{equation}
$h$ is the disorder strength, also called randomness, of the
system. Various RF probability distributions, such as the
Gaussian, the wide bimodal distribution (with a Gaussian width),
and the above bimodal distribution [equation~(\ref{eq:2})] have
been
considered~\cite{young85,gofman93,cao93,newman93,falicov95,swift97,auriac97,sourlas99,hartmann99,machta00,hartmann01,middleton02}.
As it is well known, the existence of an ordered ferromagnetic
phase for the RFIM, at low-temperature and weak-disorder, follows
from the seminal discussion of Imry and Ma~\cite{imry75}, when the
space dimension is greater than two ($D>2$). This has provided us
with a general qualitative agreement on the sketch of the phase
boundary separating the ordered ferromagnetic (\textbf{F}) phase
from the high-temperature (strong-disorder) paramagnetic
(\textbf{P}) phase. The phase boundary separates the two phases of
the model and intersects the randomness axis at the critical value
of the disorder strength, denoted hereafter as $h_{c}$. Such
qualitative sketching has been commonly used in most papers for
the RFIM~\cite{newman93,machta00,newman96,itakura01,malakis06b}
and close form quantitative expressions are also known from the
early mean-field calculations~\cite{aharony78}. However, it is
generally true that the quantitative aspects of phase diagrams
produced by mean-field treatments are very poor approximations.
This applies also for the bimodal RFIM, for which, with the
exception of the estimation of $h_{c}$ from ground-state
calculations~\cite{auriac97,sourlas99,hartmann99}, a reliable
approximation of the phase diagram is still lacking. Furthermore,
despite the $30$ years of theoretical and experimental study the
nature and scaling features of the transition of the RFIM are not
yet well understood~\cite{harris74,berker84,dotsenko07}. Nowadays,
it is generally believed that the transition from the ordered to
the disordered phase is continuous, governed by the
zero-temperature random
fixed-point~\cite{villain84,bray85,fisher86}, but a complete set
of values of the critical exponents fulfilling scaling relations
has not been established, despite the fact that several
bounds~\cite{chayes86} and further
inequalities~\cite{schwartz85,schwartz91} for the critical
exponents have been proposed, together with modified scaling
relations~\cite{nowak98}. It is also now quite clear that, the
finite-size behavior of the system is obscured by strong and
complex finite-size effects, involving the violation of
self-averaging~\cite{malakis06b,dayan93,aharony96,wiseman98,parisi02,malakis06a,fytas06,efrat07}.
In particular the issue of the order of the transition
(first-order or continuous) has regained much interest after the
recent observations of first-order-like features at the
strong-disorder regime for both the bimodal~\cite{hernandez07} and
the Gaussian RF distributions~\cite{wu05,wu06}.

This work presents a careful and systematic numerical approach to
the phase boundary of the bimodal RFIM in the low-temperature
regime. The numerical approach, presented below, is a proposal
that may be also useful to the study of other systems with complex
energy landscapes, such as general random systems, spin glasses,
proteins, and others. From our simulations, corresponding to
systems with linear sizes $L$ in the range $L=4-32$, we perform a
finite-size scaling analysis leading also to a refined value of
the critical disorder strength $h_{c}$, in good agreement with the
estimates obtained via the above mentioned ground-state
techniques. We implement a novel approach that is based on the
idea of entropic sampling in restricted energy
spaces~\cite{malakis04,malakis05} together with a reliable
extrapolation scheme and we produce accurate numerical data at the
strong-disorder regime. Our analysis of the low-temperature part
of the phase diagram provides us with a qualitative and also
quantitative description of the phase diagram of the model, also
at low values of the disorder strength, and produces good
estimates for the critical disorder strength and the correlation
length's exponent, in very good agreement with those from previous
zero-temperature studies of the model.

The rest of the paper is laid out as follows. In the next Section
we describe the numerical route implemented. In
Section~\ref{sec:3} we present in detail the low-temperature
aspects of the phase diagram of the model. Finally, we summarize
our conclusions in Section~\ref{sec:4}.

\section{Numerical Approach}
\label{sec:2}

There exist two distinct kinds of purely numerical approaches to
the RFIM. The first approach utilizes Monte Carlo methods,
including predominantly sophisticated simulation techniques, such
as cluster algorithms and flat-histogram approaches, to study
finite-temperature properties of the
system~\cite{young85,machta00,newman96,nowak98,hernandez07,hernandez97,dotsenko91,rieger93,rieger95,barber01},
while the second approach utilizes graph theoretical algorithms to
determine the ground-states and estimate the zero-temperature
behavior of the
RFIM~\cite{ogielski86,swift97,auriac97,sourlas99,hartmann99,hartmann01,middleton02,wu05,wu06,dukovski03}.
This second approach, is grounded on the belief that the critical
behavior of the model is governed by the non trivial RF
fixed-point at zero-temperature~\cite{villain84,bray85,fisher86}.

In this work, we follow a novel numerical approach by combining
current advances in simulation techniques. The proposed approach
is well adapted and efficient for the study of the RFIM at the
strong-disorder regime. Our scheme will be outlined and tested in
this Section for the 3D bimodal RFIM and it is hoped that it will
provide a convenient and fast simulation tool for studying other
similar disordered or complex systems. In effect, we shall use our
earlier idea of the entropic implementation of the WL
algorithm~\cite{malakis05}, to produce a faithful approximation of
the exchange-field two-parametric DOS of the RFIM in an
appropriate neighborhood of the disorder strength.

The WL algorithm~\cite{wang01} is one of the most refreshing
improvements in Monte Carlo simulation schemes and has been
already applied to a broad spectrum of interesting problems in
statistical mechanics and biophysics~\cite{yamaguchi01}. Several
implementations of the WL sampling technique have been carried out
by many
authors~\cite{hernandez07,wu05,wu06,yamaguchi01,okabe02,yan03,rathore04,mastny05,zhou05,malakis06c,martinos05a,malakis06d,malakis07a,malakis07b}
and the present approach may be also seen also as a further
contribution to the growing number of different applications of
the WL method in the study of complex systems with rough energy
landscapes. The original WL method has been already applied to the
RFIM in previous studies concerning the properties of the system
at specified values of the disorder strength. Such recent
investigations have been presented for the
bimodal~\cite{hernandez07} and also for the Gaussian
RFIM~\cite{wu05,wu06}, respectively. The present approach follows
the implementation of the WL random walk used already in our
recent studies of the RFIM~\cite{malakis06b,malakis06a,fytas06}.
In these studies we have carried out the WL random walk in a
restrictive and more efficient fashion. This restrictive version,
utilizes the so called critical minimum energy subspace (CrMES)
technique~\cite{malakis04,malakis05} to locate and study
finite-size anomalies of systems by carrying out the random walk
only in the dominant energy subspaces. Generally, our finite-size
scaling studies have shown that this restrictive practice can be
followed in systems undergoing
second-order~\cite{malakis04,malakis05,malakis06c,martinos05a,malakis06d}
and also first-order transitions~\cite{malakis07a,malakis07b}.
Details and tests of this approach for the 3D bimodal RFIM can
been found in reference~\cite{malakis06a}, where the thermal
properties of the system at the disorder strength value $h=2$ were
studied.

In a subsequent paper~\cite{fytas06} the magnetic properties of
the RFIM were also considered by using the same restrictive scheme
as an entropic sampling method. This simplification was introduced
and tested for the first time in our earlier work~\cite{malakis05}
on the 2D and 3D Ising models and soon after that was used for the
investigation and verification of some universal properties of the
order-parameter distribution~\cite{malakis06c}. According to this
we may estimate the magnetic properties of the systems by
recording the two-parameter energy-magnetization $(E,M)$
histograms in the final stage (high-levels) of the WL diffusion
process. At the end of the process the final accurate WL
(one-parametric) DOS $G(E)$ and the cumulative $H(E,M)$
histograms, are used to determine the magnetic properties of the
system, by forming appropriate microcanonical averages of the
order-parameter
moments~\cite{fytas06,malakis05,malakis06c,malakis06d,malakis07a,malakis07b}.

The above description may be seen as a convenient way to bypass
the requirement of a two-parametric WL sampling process and a very
similar approach will be implemented in this paper. We will now be
recording, again in the high-levels of the WL diffusion process,
the cumulative (exchange-energy, field-energy) two-parametric
histograms, in order to produce an approximation for the
two-parametric DOS of the RFIM. At this point, we should stress
that any multi-parametric WL process is inevitably restricted to
rather small
lattices~\cite{wang01,shteto97,lima00,silva06,tsai07}. In fact the
applications of such multi-parametric methods are substantially
limited, since besides the immense time and excessive memory
requirements, they very often face severe ergodic and/or
convergence problems, depending on both the physical system and
the algorithmic implementation. However, notable examples of such
two-parametric studies, mainly on 2D systems, discussing also some
of the above problems, have been carried out in the last $10$
years. The most recent two-parametric investigation performed by
Tsai et al.~\cite{tsai07} concerns the critical endpoint of the 2D
asymmetric Ising model with two and three-body interactions on the
triangular lattice. This last study required several days of
computer time and a quite large computer memory for the larger
lattice size studied, consisting of $N=42\times 42$ lattice
points. To our knowledge, this is also the largest system that has
been reported by the two-parametric WL algorithm. Certainly, a
similar two-parametric study is possible, although lacking, for
the RFIM. However, the correspondingly large 3D system will have
linear sizes of the order of $L=12$, and this will be very small
for our purposes. It will be seen in the next Section, that such
lattice sizes are rather small for an accurate estimation of
$h_{c}$ of the bimodal RFIM.

We now proceed to give the details of the present entropic
implementation of the WL approach. Carrying out the WL process at
a particular value $h$ of the disorder strength, we attempt to
generate good approximations of the (exchange-energy,
field-energy) two-parametric DOS for the RFIM in a neighborhood of
$h$. An analogous approach was undertaken several years ago,
before the invent of the WL method, by Deserno~\cite{deserno97},
who used flat-histogram techniques and also a restricted energy
sampling to locate and study some properties of the tricritical
point of the Blume-Capel model~\cite{blume66} on a simple cubic
lattice. The extrapolation scheme described below subjects to the
following WL process: depending on the lattice size, we use a
total of at least $j_{WL}=20$ WL iterations, producing at each
iteration level well-saturated energy-histogram
fluctuations~\cite{lee06} and obeying at least the $5\%$ flatness
criterion~\cite{malakis04,malakis05}. The reduction of the WL
modification factor follows the usual rule:
$f_{j+1}=\sqrt{f_{j}}$,
$f_{1}=e$~\cite{malakis04,malakis05,wang01}, and the range
$j_{WL}\geq 16$ of the WL process is used for the recording and
accumulation of the appropriate energy histograms (see definitions
below).

To introduce our notation, let us now conveniently separate the
Hamiltonian of equation~(\ref{eq:1}) of the RFIM as follows
\begin{equation}
\label{eq:3}
\mathcal{H}(x)=-J\mathcal{H}_{J}(x)-h\mathcal{H}_{h}(x)=-\mathcal{H}_{J}(x)-h\mathcal{H}_{h}(x),
\end{equation}
where $x$ denotes a spin state in phase space and we have set
$J=1$, since the behavior of the model depends only on the ration
$h/J$. Assuming that the two-dimensional DOS $G(E_{J},E_{h})$ in
the exchange and field variables $E_{J}=\mathcal{H}_{J}(x)$ and
$E_{h}=\mathcal{H}_{h}(x)$ is known, the DOS with respect to the
total energy $E=\mathcal{H}(x)=-E_{J}-h'E_{h}$ corresponding to
any value $h'$ of the disorder strength, can be deduced by summing
over all pairs giving the particular value of the total energy
\begin{equation}
\label{eq:4} G_{h'}(E)=\sum_{E_{J}+h'E_{h}=E} G(E_{J},E_{h}).
\end{equation}
Let us further assume an entropic Markov process in which $M$ spin
states are selected from the phase space with probability
$w_{h}(x)$ depending on the DOS $G_{h}(E)$, where $E$ is the total
energy of the spin state at the value $h$ of the disorder,
\begin{equation}
\label{eq:5} w_{h}(x)\propto [G_{h}(E)]^{-1}.
\end{equation}
Then, an approximation of the two-parametric (exchange-energy,
field-energy) DOS of the RFIM in a neighborhood of $h$ is provided
by the expectation of the observable
$\delta_{E_{J};\mathcal{H}_{J}}\delta_{E_{h};\mathcal{H}_{h}}$
\begin{eqnarray}
\label{eq:6} \tilde{G}^{(h)}(E_{J},E_{h})&\simeq&
\frac{1}{Mw_{h}(x)}\sum_{x\in
\{x\}_{M}}\delta_{E_{J};\mathcal{H}_{J}}\delta_{E_{h};\mathcal{H}_{h}}\nonumber
\\ &\simeq& G_{h}(E)\frac{H^{(h)}(E_{J},E_{h})}{H^{(h)}(E)},
\end{eqnarray}
where the last equality follows from equation~(\ref{eq:4}), using
the above approximate two-dimensional DOS in place of the exact
and observing that $H^{(h)}(E)=\sum_{E_{J}+hE_{h}=E}
H^{(h)}(E_{J},E_{h})$ and the double histogram
$H^{(h)}(E_{J},E_{h})$ is the above sum of the observable
$\delta_{E_{J};\mathcal{H}_{J}}\delta_{E_{h};\mathcal{H}_{h}}$.
The superscript $(h)$ in the quantities in the above equation is
only a reminder of the fact that the simulation is performed at
the disorder strength value $h$. It should be noted that this
notation does not mean an $h$-dependence, but rather a statistical
indirect influence on the reliability of the histogram recordings
and accordingly on the two-dimensional DOS. In our approach the
ratio of histograms in the above equation~(\ref{eq:6}), by the
assumed Markov process, is replaced by the ratio developed during
the final high-levels ($j_{WL}\geq 16$) of the WL process.
Denoting these latter histograms by $H^{(h)}_{WL}(E_{J},E_{h})$
and $H^{(h)}_{WL}(E)$ and by $\tilde{G}_{h}(E)$ the WL DOS, as
modified at the final level of the process, our final
approximation reads
\begin{equation}
\label{eq:7} \tilde{G}_{WL}(E_{J},E_{h})\simeq
\tilde{G}_{h}(E)\frac{H^{(h)}_{WL}(E_{J},E_{h})}{H^{(h)}_{WL}(E)}.
\end{equation}

The above approximation provides in conjunction with the skew
summing procedure of equation~(\ref{eq:4}) a suitable
extrapolation scheme which can be used for the study of the RFIM
in the neighborhood of the disorder value $h$, used for the WL
simulation. This extrapolation program will be applied in the next
Section for the study of the finite-size development of the phase
diagram of the bimodal RFIM at the strong-disorder regime. From
our previous studies it has been
verified~\cite{malakis05,malakis06c} that the detailed balance
condition is quite well satisfied at the high-levels of the WL
process and the recording of appropriate histograms produces
faithful and good approximations. Therefore, it is hoped that the
proposed extrapolation program will not produce systematic errors,
besides the expected statistical fluctuations. However, for safety
reasons, we shall use relatively small values for the
extrapolation shift parameter $|h-h'|$, at most of the order of
$7\%$ of the disorder strength value, and a rather loose
restriction of the energy space in which the main WL simulation is
performed. In particular, in most of our simulations performed at
$h=2.25$ the energy spectrum for the simulation was restricted
only from the high-energy side, while the entire low-energy part
of the spectrum down to the ground-state was included (see also
the discussion below). For the restriction of the high-energy side
we used our data from our previous study of the model at the value
$h=2$. In this way the WL sampling extends to all energy values
with a significant contribution to the finite-size anomalies of
the model for all values $h>2$. For moderately large lattice sizes
($L>12$), this practice is not the optimum choice. This is
because, besides the energy states contributing to the range
$h=2.1-2.4$, used in our extrapolation program, one simulates also
the part of the energy spectrum in the neighborhood of the
ground-states in which the convergence of the algorithm is very
slow. Thus, for the larger lattice sizes, one may avoid this
ground-state neighborhood, as we have done for the sizes $L=26$
and $L=32$.
\begin{figure}
\resizebox{1 \columnwidth}{!}{\includegraphics{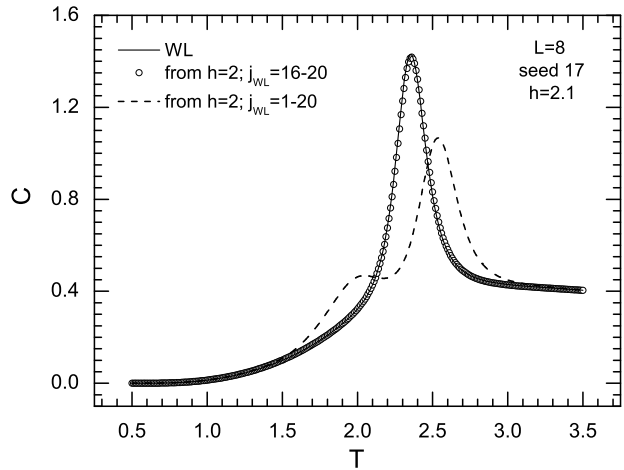}}
\caption{Illustration of the effect of the violation of the
detailed balance in the early WL iteration levels. Details of the
shown approximations are also given in the text.} \label{fig:1}
\end{figure}

Before moving to the presentation of our results, let us end this
Section by presenting some tests on the reliability of the
proposed approach. Figure~\ref{fig:1} illustrates the accuracy of
our practice of using the high-levels of the WL process as an
entropic sampling method. The curves and points shown represent
three different approximations of the specific heat for a
particular RF on a lattice of linear size $L=8$. The solid line is
the directly simulated specific heat by the WL method at $h=2.1$
and should be seen as an almost exact result. The open circle
points represent an excellent approximation obtained for the value
$h=2.1$ by using a WL simulation at $h=2$ and our extrapolation
scheme, using the high-WL iteration levels ($j_{WL}=16-20$) for
the recording the double (exchange-energy, field-energy)
histograms. Finally, the dashed line shows some quite dramatic
distortion effects obtained by using the whole ($j_{WL}=1-20$) WL
iteration range for the recording of the above two-dimensional
energy histograms. This is of course an example, showing possible
subtle effects coming from a significant violation of the detailed
balance condition in the early WL iteration levels.

\begin{figure}
\resizebox{1 \columnwidth}{!}{\includegraphics{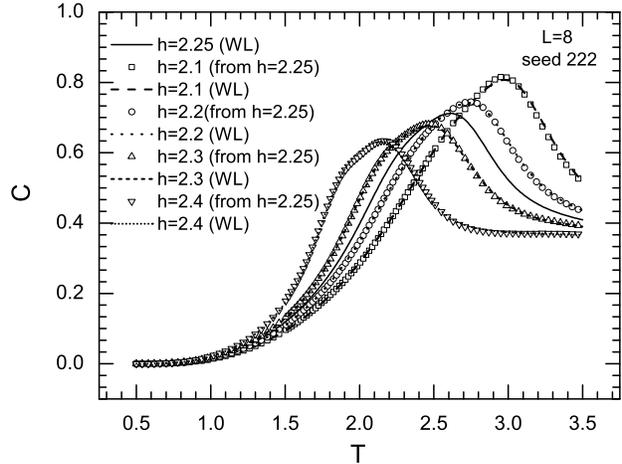}}
\caption{Specific heat curves for a certain RF of the lattice size
$L=8$. Illustration of the behavior for several values of the
disorder strength obtained by direct WL simulation (lines) and by
the extrapolation scheme (points).} \label{fig:2}
\end{figure}
\begin{figure}
\resizebox{1 \columnwidth}{!}{\includegraphics{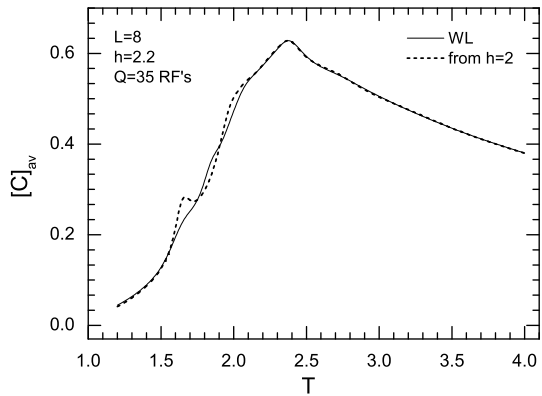}}
\caption{Average specific heat curve at $h=2.2$, obtained by
direct WL simulation (solid line) and by extrapolation (dotted
line), for lattice size $L=8$ averaged over $Q=35$ RF's.}
\label{fig:3}
\end{figure}

A second test showing now the reliability of our extrapolation
scheme is presented in Figure~\ref{fig:2}. Here we show specific
heat curves, in the range $h=2.1-2.4$, obtained by the proposed
extrapolation scheme from a WL simulation performed at $h=2.25$,
together with the results obtained independently via direct WL
simulation at the corresponding disorder strength values. For
values very close to $h=2.25$, the two different approximations
coincide, and even for the values $h=2.1$ and $h=2.4$ there are
only very small deviations, mainly around the peaks. The locations
of the pseudocritical temperatures are very weakly dependent on
the extrapolation scheme and are therefore quite accurately
determined by the method. The effects on ensemble averages will be
expected to be even weaker. This is illustrated in our final test
concerning the pseudocritical temperatures obtained from the
ensemble average specific heat curve, used in the next Section for
the description of the phase diagram. The average specific heat is
defined as usually~\cite{rieger93,rieger95}
\begin{equation}
\label{eq:8} [C]_{av}=\frac{1}{Q}\sum_{q=1}^{Q}C_{q}(T),
\end{equation}
where the index $q=1,\ldots,Q$ runs over the number of disorder
realizations. Figure~\ref{fig:3} concludes this Section by a
comparison of two approximations of the average specific heat
curve $[C]_{av}$ obtained from an ensemble of $Q=35$ realizations
and corresponding to the disorder strength value $h=2.2$. The
solid line is the average curve obtained by a direct WL simulation
at $h=2.2$, while the dashed line represents the approximation of
the extrapolation scheme based on a WL simulation on the same
ensemble at the value $h=2$. Clearly, the locations of the two
pseudocritical temperatures coincide and the two specific heat
peaks are in excellent agreement.

\section{Phase Diagram}
\label{sec:3}

We aim here to present a reliable approximation of the phase
diagram of the 3D bimodal RFIM at the strong-disorder regime and
provide an accurate estimate for $h_{c}$ (independent from the
ground-state approach). Despite the general qualitative agreement
between different approaches on the phase diagram of the model,
the various estimations throughout the literature vary in a rather
wide range. This diversity on the numerical estimation of the
phase diagram is true for both the Gaussian and the bimodal
distributions and is generally reflected in the wide range of
estimates for $h_{c}$. Thus, for the Gaussian RFIM the values for
$h_{c}$ span the range $h_{c}=2.26-2.37$, despite the fact that
these values are mainly estimated via the same ground-state
approach~\cite{ogielski86,swift97,auriac97,hartmann99,hartmann01,middleton02,newman96,nowak98,wu06,dukovski03}.
On the other hand, there are fewer attempts devoted to the
estimation of the phase diagram of the bimodal RFIM and the
corresponding estimates for $h_{c}$, obtained again via the
ground-state approach, are restricted in a smaller range, i.e.
$h_{c}=2.20-2.25$~\cite{swift97,auriac97,hartmann99}. Our previous
attempt to estimate the phase diagram using a high-temperature
(weak-disorder: $h=0.5-2$) numerical study yielded an
overestimation for $h_{c}$, namely
$h_{c}=2.42(18)$~\cite{malakis06b}. However, we will show here,
that an accurate estimation of the phase diagram is possible by a
more systematic low-temperature (strong-disorder: $h\geq 2$)
numerical study. In this case, we will find a much lower estimate
for $h_{c}$ that agrees favorably with the estimates given above
from the ground-state approach. Additional good comparisons with
some phase diagram points, estimated earlier in the literature,
provide evidence that our final proposal for the phase diagram may
be a reliable and competent approximation for the whole disorder
strength range.

We proceed here to analyze our numerical data at the
strong-disorder regime. Using our entropic implementation of the
WL method and the extrapolation procedure, outlined in the
previous Section, we have generated numerical data for the
following lattice sizes: $L\in\{4,8,12,16,20,26,32\}$. For lattice
sizes in the range $L=4-20$ we have simulated $20$ RF's, whereas
for the larger sizes $L=26$ and $L=32$, $10$ realizations of the
RF have been simulated. For each lattice size and each
realization, we performed a WL simulation in an appropriate energy
subspace, restricted only from the high-energy end and including
the entire low-energy spectrum down to the ground-state, with the
exception of the sizes $L=26$ and $L=32$ for which the very close
to the ground-sate energy levels were avoided. The WL simulation
was performed at the disorder strength value $h=2.25$ and the
accumulated double (exchange-energy, field-energy) histogram was
then used to approximate the two-parametric DOS
[equation~(\ref{eq:7})] and finally, the DOS $G_{h'}(E)$
[equation~(\ref{eq:4})] and the thermal properties of the system
for various values of randomness in a neighborhood of the
simulated value $h=2.25$. In order to construct the average
specific heat curve (Figure~\ref{fig:3}) and to identify via its
peak a pseudocritical temperature $T_{L;h}$, representing the
ensemble of RF's at the particular lattice size, a summation over
the realizations was performed, as in equation~(\ref{eq:8}). As
discussed earlier and illustrated in Figures~\ref{fig:1} -
\ref{fig:3}, the described extrapolation scheme provides a
reliable approximation of the location of the maximum of the
average specific heat curve. The systematic shift of the
individual specific heat peaks, shown in Figure~\ref{fig:2}, for
higher values of $h$, will be reflected in the corresponding
shifts of the peaks of the average specific heat curves, as should
be expected, providing us the necessary information for the
finite-size analysis. The locations of all these specific heat
peaks, for all lattice sizes mentioned above, were calculated from
our simulation data at $h=2.25$, and their extrapolations to other
neighbor $h$-values, for the following set I of disorder values,
set I: $h'=\{2.1,2.15,2.2,2.25,2.3,2.35,2.4\}$. For the lattice
size $L=12$, an additional entropic WL sampling was carried out at
$h=2$, using now a larger ensemble of $Q=250$ RF's. Again, using
the extrapolation procedure of equations~(\ref{eq:7}) and
(\ref{eq:4}) the specific heat peaks corresponding to the
following set II of disorder values were located, set II:
$h'=\{1.7,1.8,1.9,2,2.1,2.2\}$.
\begin{figure}
\resizebox{1 \columnwidth}{!}{\includegraphics{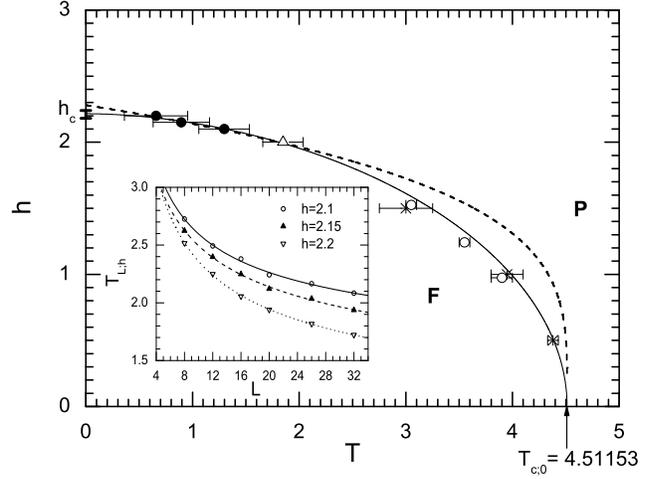}}
\caption{Approximations of the phase diagram of the 3D bimodal
RFIM. Two fitting attempts are shown. The solid line corresponds
to the elliptical ansatz (\ref{eq:10}) giving $h_{c}=2.215(35)$,
while the dashed line to the power-law ansatz (\ref{eq:11}) giving
$h_{c}=2.277(49)$. The range of ground-state estimates for $h_{c}$
and the zero-field's critical temperature $T_{c;0}=4.51153$ are
marked on the axis. The inset shows the shifting of the
pseudocritical temperature $T_{L;h}$ for three values of the
disorder strength, i.e. $h=2.1$, $2.15$, and $2.2$.} \label{fig:4}
\end{figure}

Let us attempt now a finite-size analysis using the size-shifts of
the pseudocritical temperatures of the averaged specific heat
curves for some particular value of the disorder. The inset of
Figure~\ref{fig:4} illustrates fitting attempts of these
size-shifts for three values of the disorder. The range $L=8-32$
is used in these fits by assuming the usual power law:
\begin{equation}
\label{eq:9} T_{L;h}=T_{c;h}+a_{h}L^{-1/\nu_{h}}.
\end{equation}
The critical temperatures $T_{c;h}$, resulting as limiting values
of the corresponding pseudocritical temperatures, for the attempts
shown in the inset of Figure~\ref{fig:4}, are $1.297(237)$,
$0.894(264)$, and $0.659(299)$, for the disorder strengths
$h=2.1$, $2.15$, and $2.2$ respectively. We have excluded, from
our fitting attempts here, the lattice size $L=4$ in order to
eliminate the influence from the very small $L$-behavior and this
practice will be followed and further discussed in the sequel.
Following the same fitting procedure, again in the range $L=8-32$,
for $h=2.25$ we find that the corresponding critical temperature
becomes now negative, i.e. $T_{c;2.25}=-0.18$. This fact shows
that, within our fitting scheme, the value of the disorder
strength $h=2.25$ is an upper bound for the critical disorder
strength. Noteworthy, that if we use the range $L=4-32$ instead,
the negative sign for the critical temperature will appear at the
value $h=2.35$, which however appears to be a rather
overestimating bound for the critical randomness. Thus, only the
three points $h=2.1$, $h=2.15$, and $h=2.2$ (filled circles)
resulting from the fits shown in the inset of Figure~\ref{fig:4}
can be used to approximate the phase diagram. In order to find one
more point of the phase diagram we shall now also use our earlier
numerical data~\cite{malakis06a} (from rather large $Q=500-1000$
ensembles of RF's) for the disorder strength $h=2$. Using the
above fitting practice in the range $L=8-32$ we find from the
general pseudocritical temperature shift behavior the limiting
value $T_{c;2}=1.848(188)$ (open triangle), which is just inside
the estimate bounds given in our previous paper
($T_{c;2}=2.03(18)$) using sizes in the range
$L=4-32$~\cite{malakis06a}.

The above four approximate phase diagram points, corresponding to
the disorder strength values $h=2$, $2.1$, $2.15$, and $2.2$, will
be now used to find a phenomenological representation of the phase
diagram of the bimodal RFIM. Let us first attempt an elliptical
fit using the following ansatz
\begin{equation}
\label{eq:10}
h=h_{c}\sqrt{1-\left(\frac{T_{c;h}}{\tau}\right)^{x}}.
\end{equation}
The rescaling temperature factor $\tau$ in equation~(\ref{eq:10})
will be handled either as a free-parameter during the fit, or as a
fixed-parameter using the best known estimate for the critical
temperature of the zero-field Ising model, namely
$T_{c;0}=4.51153$~\cite{ferrenberg91}. The resulting phase
diagrams almost coincide (see Figure~\ref{fig:4} where for clarity
reasons only the latter case is shown) and are described
respectively by the following ($h_{c}$, $\tau$, $x$; $\chi^{2}$)
parameter values, including the value of the $\chi^{2}$-test:
($2.212(29)$, $4.50394(778)$, $1.862(87)$; $\sim 10^{-4}$) and
($2.215(35)$, $4.51153$, $1.847(92)$; $\sim 10^{-4}$),
respectively. Thus, our fitting attempts with
equation~(\ref{eq:10}) produce a value for the critical disorder
which is very close to the estimates obtained from the
zero-temperature studies of the
model~\cite{swift97,auriac97,hartmann99}. Furthermore, the fitting
using the temperature rescaling factor $\tau$ in
equation~(\ref{eq:10}) as a free-parameter produces a fairly good
estimate for the critical temperature of the zero-field Ising
model~\cite{ferrenberg91}.

As an alternative to the above elliptical fit, we have also
considered for comparison the following power-law
ansatz~\cite{malakis06b}
\begin{equation}
\label{eq:11}
h=h_{c}\left(\frac{T_{c;0}-T_{c;h}}{T_{c;0}}\right)^{x}.
\end{equation}
The attempt to fit the same data to this law is illustrated also
in Figure~\ref{fig:4} by the dashed line. In this case we find a
noticeable overestimation of $h_{c}$, namely $h_{c}=2.277(49)$ and
a much larger (by a factor of $70$) value of $\chi^{2}$ of the
fit. Therefore, we conclude that the elliptical law of
equation~(\ref{eq:10}) provides a better representation of the
phase diagram of the RFIM. Of course, our attempt above aims only
at a numerical approximation for the main part of the diagram and
not at the correct asymptotic behavior at its ends. For instance,
the behavior of the phase diagram at a very small neighborhood
around the critical temperature of the pure system, is expected to
be determined by the susceptibility exponent $\gamma$ of the pure
system~\cite{aharony78,berker84,andelman84}, as follows from the
phenomenological renormalization arguments of
reference~\cite{berker84}. Accordingly, the slope of the phase
diagram at this end is expected to behave as $\delta h \sim
(\delta T)^{\gamma}$ (where $\gamma=1.2358$ for the pure 3D Ising
model~\cite{talapov96}) and not with the exponent $1/2$ of the
ansatz (\ref{eq:10}). It appears that similar elliptical laws have
been also used previously by other authors for the Gaussian
RFIM~\cite{machta00,newman96}, although these were not stated
explicitly.

Finally, we would like to note that we have included in
Figure~\ref{fig:4} some more data points for smaller values of the
disorder strength from previous numerical works. These are the
data for $h=0.5$, $1$, and $1.5$ (shown by stars in the figure)
from our previous investigation of the phase diagram of the
model~\cite{malakis06b} and three more points (open circles)
estimated by Rieger and Young~\cite{rieger93}. These points are
close enough to our approximate phase diagram and the small
deviation comes possibly from the fact that these have been
estimated, in both cases, by applying equation~(\ref{eq:9}) to
rather small sizes: $L\leq 24$ and $L\leq 16$, respectively (see
also the discussion below).

At this point, let us comment on the significance of our notation
concerning the shift exponent $\nu_{h}$ in equation~(\ref{eq:9}).
As mentioned earlier, we have tried to avoid the influence of the
very small $L$-behavior in our estimates, thus excluding from our
fitting attempts the data for $L=4$. This is a compromise followed
because in our study (and in effect in all finite-temperature
studies) a rather restricted $L$-range is available for performing
finite-size scaling analysis. However, it has been pointed out in
reference~\cite{malakis06b} that the estimates based on such
restricted ranges should not be completely trusted and this may be
particularly true for the shift exponent $\nu_{h}$. For instance,
the range $L=4-24$ will produce quite different estimates, for
both $T_{c;h}$ and $\nu_{h}$, from those obtained above from the
range $L=8-32$ and this fact, together with the use of the
power-law in equation~(\ref{eq:11}), are the two reasons behind
our earlier overestimation of $h_{c}$ ($h_{c}=2.42(18)$ in
reference~\cite{malakis06b}). The general asymptotic behavior of
the RFIM follows different complex routes that appear to strongly
depend on the value of the disorder strength $h$ and different
ranges of lattice sizes may be needed in order to approach the
asymptotic behavior for different values of disorder strengths.
Even the observation of an appreciable disorder strength
dependence on $\nu_{h}$, should be reluctantly identified as a
possible violation of universality along the phase boundary,
although this violation of universality is one of the strongly
supported scenarios in the literature~\cite{sourlas99}. The
violation of universality for the case of the 3D RFIM has been
discussed a few years ago by Sourlas~\cite{sourlas99}. Equivalent
studies of universality violations have been reported also in
other glassy systems~\cite{bernardi96}, reenforcing the view that
the concept of universality in complex systems is not fully
clarified.
\begin{figure}
\resizebox{1 \columnwidth}{!}{\includegraphics{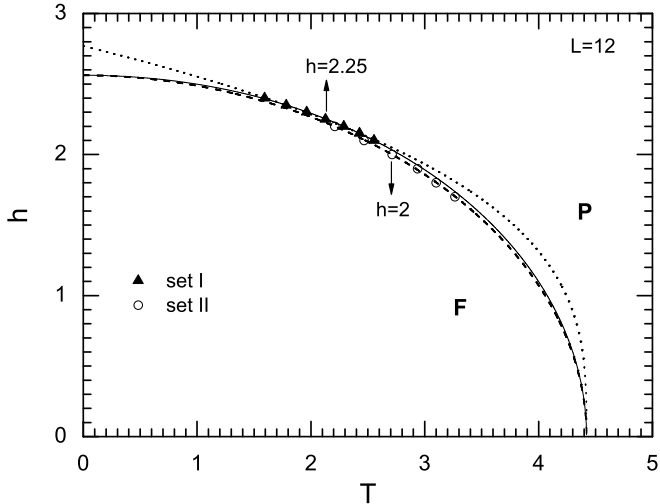}}
\caption{Finite-size elliptical phase diagrams for $L=12$ using
two different extrapolation sets of disorder strength values (set
I (filled triangles) and set II (open circles)) and different
realizations ensembles. The solid and dashed lines are elliptical
fits of the form (\ref{eq:12}) with comparable values of
$\chi^{2}$ of the order of $10^{-7}$ giving for the pseudocritical
disorder strength the values $h_{c;12}=2.56(3)$ and
$h_{c;12}=2.56(2)$, respectively. The application of the
finite-size version of the power-law (\ref{eq:11}), shown by the
dotted line, has a larger value of $\chi^{2}=10^{-5}$ and produces
an overestimation for the pseudocritical disorder strength:
$h_{c;12}=2.77(5)$.} \label{fig:5}
\end{figure}

We proceed now with an alternative estimation of the critical
disorder strength. Firstly, let us point out that for each value
of $L$, our data can be used to produce a finite-size phase
diagram. Provided that the phase diagram points do not decline
appreciably from the above elliptical law, we may attempt to
construct a finite-size sequence of diagrams by using the
finite-size version of equation~(\ref{eq:10})
\begin{equation}
\label{eq:12}
h=h_{c;L}\sqrt{1-\left(\frac{T_{L;h}}{\tau_{L}}\right)^{x}},
\end{equation}
where now the rescaling temperature factor $\tau_{L}$ may be
either handled as a free-parameter during the fit or as a
fixed-parameter at the corresponding zero-field's Ising model
pseudocritical temperatures taken from Table IV of
reference~\cite{malakis04}. Using this latter choice for
$\tau_{L}$, Figure~\ref{fig:5} provides a test of this approach
producing two very similar phase diagrams for the size $L=12$. The
two diagrams are obtained using the two different sets of phase
diagram points corresponding to set I and set II of the disorder
strength values. The first set of points (filled triangles) is
determined over an ensemble of $Q=20$ realizations of the RF and
corresponds to set I, i.e. simulation at $h=2.25$ and suitable
extrapolation in the range $h=2.1-2.4$. The other set of points
(open circles) is determined over a larger ensemble of $Q=250$
realizations of the RF and corresponds to set II, i.e simulation
at $h=2$ and extrapolation in the range $h=1.7-2.2$. The
application of the elliptical law (\ref{eq:12}) gives the two very
similar phase diagrams shown in Figure~\ref{fig:5} by the solid
and dashed lines for the two set of points, respectively. These
two diagrams, with comparable values for $\chi^{2}$ of the order
of $10^{-7}$, come together to the same point at $T=0$, giving the
value $h_{c;12}=2.56$ for the finite-size ($L=12$) critical
disorder strength. For illustration reasons, we have also included
in this figure the attempt using the corresponding finite-size
version of equation~(\ref{eq:11}) for the set of points obtained
from the simulations at the value $h=2.25$ (dotted line). Again
the $\chi^{2}$ quality of the fit is worst for the power-law
($\chi^{2}=10^{-5}$) and produces a clear overestimation for the
pseudocritical disorder strength of the order of
$h_{c;12}=2.77(5)$. The comparison between the two finite-size
elliptical phase diagrams, corresponding to the two sets of points
($h=2.25$ and $h=2$), is on the other hand very convincing. Thus,
Figure~\ref{fig:5} provides strong evidence in favor of our choice
of using in our simulations for all lattice sizes the
strong-disorder regime corresponding to the value $h=2.25$. In
particular it shows that our data based on only the $Q=20$ RF's
are sufficient for our proposes of estimating the phase diagram.

\begin{figure}
\resizebox{1 \columnwidth}{!}{\includegraphics{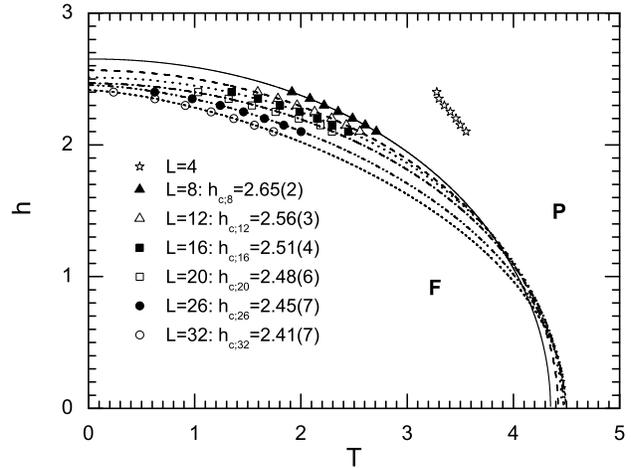}}
\caption{Finite-size elliptical phase diagrams for lattice sizes
in the range $L=8-32$, using set I of the disorder strength
values. The drawn lines represent the finite-size elliptical
fittings according to equation~(\ref{eq:12}), in which the
rescaling temperature factor $\tau_{L}$ was fixed at the
corresponding zero-field's Ising model pseudocritical
temperatures.} \label{fig:6}
\end{figure}
\begin{figure}
\resizebox{1 \columnwidth}{!}{\includegraphics{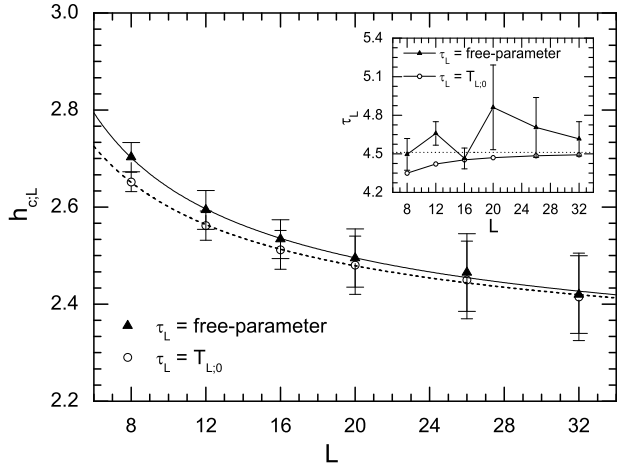}}
\caption{Shift behavior of the finite-size critical disorder
strengths $h_{c;L}$. The inset shows the oscillations in the
values of $\tau_{L}$, when this parameter is handled as a
free-parameter. Their behavior follows the correct trend,
approaching the zero-field's $T_{c;0}=4.51153$~\cite{ferrenberg91}
(dotted line).} \label{fig:7}
\end{figure}

Figure~\ref{fig:6} presents the finite-size elliptical phase
diagrams for lattice sizes in the range $L=8-32$, using set I of
the disorder strength values. For the lattice size $L=4$ we have
not drawn a finite-size phase diagram, since it is quite evident
from the corresponding open star points in Figure~\ref{fig:6} that
they decline very early, at about the value $h=2.2$, from the
elliptical law. No such deviation is observed for the other
lattice sizes, within the set I of disorder values, and this
fortifies our choice to use the particular set I for these lattice
sizes. Of course, an attempt to push our approach to even larger
lattices may require a WL simulation at $h=2.2$ and a
corresponding set of somewhat smaller disorder values. The drawn
lines in Figure~\ref{fig:6} represent the finite-size elliptical
fittings according to equation~(\ref{eq:12}), in which the
rescaling temperature factor $\tau_{L}$ was fixed at the
corresponding zero-field's Ising model pseudocritical
temperatures. For clarity the diagrams using $\tau_{L}$ as a
free-parameter are not shown. However, the main frame and the
inset of Figure~\ref{fig:7} illustrate the smoothness of the both
fitting schemes and reveal a convincing and regular shift-behavior
of the finite-size critical disorder strengths $h_{c;L}$. This
behavior allows now a finite-size analysis for the estimation of
$h_{c}$. The solid and dashed lines show good quality fits to the
following usual shift power-law
\begin{equation}
\label{eq:13} h_{c;L}=h_{c}+bL^{-1/\nu}.
\end{equation}
Thus, the fitting attempts in Figure~\ref{fig:7} produce estimates
for the asymptotic value of the critical disorder strength
$h_{c}$, and the corresponding shift-exponent $\nu$. The fitting
scheme based on the estimates of the pseudocritical disorder
strengths $h_{c;L}$ (open circles in Figure~\ref{fig:7}), produced
by fixing the rescaling temperature factor $\tau_{L}$ at the
corresponding zero-field's Ising model pseudocritical
temperatures, i.e. $\tau_{L}=T_{L;0}$, gives $h_{c}=2.219(83)$ and
$\nu=1.806(390)$. Finally, the fitting attempt based on the
corresponding $h_{c;L}$ estimates (filled triangles in
Figure~\ref{fig:7}), produced by using $\tau_{L}$ as a
free-parameter, results in a almost identical estimate for the
critical disorder, i.e. $h_{c}=2.219(65)$ but a slightly lower
estimate for the shift-exponent $\nu=1.640(423)$.  From the inset
of Figure~\ref{fig:7} we may note some oscillations in the values
of $\tau_{L}$, when this is handled as a free-parameter, which
however appear to follow the correct trend so that $\tau_{L}$ will
approach $T_{c;0}$ (dotted line) with increasing lattice size. In
both cases, the estimates for $h_{c}$ compare very well with those
obtained above from fitting equation~(\ref{eq:10}) in
Figure~\ref{fig:4} and also with those of the ground-state
approach~\cite{swift97,auriac97,hartmann99}. Despite the deviation
of the two estimates for the shift-exponent and the relatively
very large variation of $\nu$ in the literature (for both the
Gaussian and bimodal cases), it is of interest to compare here the
estimate of the second case ($\nu=1.64$) with the estimates
$1.67(11)$ and $1.66(8)$ of references~\cite{auriac97} and
\cite{hartmann99} respectively, obtained by zero-temperature
simulations.

The above observations provide concrete evidence in favor of our
present approach. It appears that, this method may be capable to
produce, if further pushed to larger lattices, even more accurate
estimates for both the critical disorder strength and also the
$T=0$ correlation length exponent, assuming that its behavior
follows the observed shift-behavior of our finite-size projections
$h_{c;L}$. It is well known from the general scaling theory that,
even for simple models, the equality between the correlation
length's exponent and the shift exponent is not a necessary
consequence of scaling~\cite{barber83}. Of course, it is a general
practice to assume that the correlation length behavior can be
deduced by the shift behavior of appropriate thermodynamic
functions. In our view, the recent strong version of the
zero-temperature fixed-point scenario by Wu and
Machta~\cite{wu05,wu06}, supports the above assumption that the
finite-size projections $h_{c;L}$ are appropriate shifting
parameters. The thermal states of Wu and Machta (see Figure 4 of
reference~\cite{wu05}) at temperatures close to the finite-size
anomalies are strongly correlated to the ground-states at disorder
strength values close to the zero-temperature critical point and
this strong correlation may be seen as a phenomenological
justification of our assumption.

\section{Conclusions}
\label{sec:4}

A numerical approach combining well-known techniques has been
proposed as a convenient alternative for the study of disordered
systems. Within this approach, the well-known WL algorithm is
used, at its final stage, as an entropic sampling method, and
multi-parametric histograms, appropriate for the study of the
system, are produced. The main advantage of this scheme is that
the requirement of multi-parametric WL sampling is surpassed and
by using the DOS, obtained via the WL method, and the accumulated
histogram information, the thermal properties of the disordered
system may be obtained in a neighborhood of the simulated disorder
strength value. The numerical techniques presented in this paper
may find further applications in the study of critical properties
of other challenging disordered systems. Via the above approach,
we have studied the general size-shift behavior of the
low-temperature part of the phase diagram of the 3D bimodal RFIM.
Our detailed analysis provided an overall reliable representation
of the main part of the phase diagram, yielding accurate estimates
for the critical disorder strength. These estimates are in
agreement with those from previous zero-temperature studies of the
model including the estimates for the correlation length's
exponent.

As a closing remark, we would like to mention that, using our WL
DOS's - for some typical RF realizations, at the simulated
disorder strength value $h=2.25$ - we have also observed, for the
larger sizes studied, first-order-like double peaks in the energy
probability densities, in agreement with the recent observations
of Hern\'{a}ndez and Ceva~\cite{hernandez07}, and Wu and
Machta~\cite{wu05,wu06}, mentioned in the introduction. This main
issue appears to be still a matter of controversy and we are
currently carrying out further research in order to clarify the
persistence (or not) of such first-order-like characteristics in
the asymptotic limit. However, the full resolution of this aspect
requires an understanding of the complex finite-size effects of
the RFIM at the strong-disorder regime and substantial computer
resources to be devoted for the simulation of large ensembles of
RF realizations in a convenient neighborhood of disorder strength
values.

\begin{acknowledgement}
The authors would like to thank Professor A.N. Berker for useful
discussions. This research was supported by the Special Account
for Research Grants of the University of Athens under Grant No.
70/4/4071. N.G. Fytas acknowledges financial support by the
Alexander S. Onassis Public Benefit Foundation.
\end{acknowledgement}

\end{document}